%% file: EASE-I2I-2018.tex
\documentclass[sigconf]{acmart}
\usepackage{fancybox}
\usepackage[export]{adjustbox}

\usepackage{pifont} 
\usepackage{subfig}
\usepackage{colortbl}
\usepackage{hhline} 
\usepackage{enumitem}
\usepackage{tikz}

\usetikzlibrary{patterns}
\tikzset{ 
table/.style={
  matrix of nodes,
  row sep=-\pgflinewidth,
  column sep=-\pgflinewidth,
  nodes={draw,rectangle,text width=2cm,align=center},
  text depth=1.5ex,
  text height=4.5ex,
  nodes in empty cells
}
}

\tikzset{
    every picture/.style={
        remember picture,   
        inner xsep=0pt, 
        inner ysep=1pt, 
        baseline,       
        every node/.style={
            anchor=base 
        }
    }
}

\usepackage{booktabs} 

\usepackage{listings}
\usepackage{color}

\definecolor{dkgreen}{rgb}{0,0.5,0}
\definecolor{gray}{RGB}{224,224,224}
\definecolor{mauve}{rgb}{0.58,0,0.82}
\definecolor{lb}{RGB}{204,255,229}

\newcommand{\cm}[1]{}

\usepackage{framed}
\usepackage{graphicx,multirow}

\setcopyright{rightsretained}
\usepackage{booktabs} 
\newcommand{\ignore}[1]{}
\usepackage{xcolor,colortbl}
\definecolor{G1}{RGB}{249, 246, 246}
\definecolor{light}{RGB}{189,213,215}
\definecolor{blue1}{RGB}{153,205,255}
\definecolor{g1}{RGB}{195, 207, 222}
\definecolor{blue2}{RGB}{0,128,255}
\definecolor{blue3}{RGB}{0,76,153}
\definecolor{red1}{RGB}{255,180,180}
\definecolor{red2}{RGB}{255,95,95}
\definecolor{red3}{RGB}{205,0,0}
\definecolor{green}{RGB}{229,255,204}
\definecolor{red}{rgb}{0.85,0.8,0.9}			

\usepackage{etoolbox}
\makeatletter
\patchcmd{\maketitle}{\@copyrightspace}{}{}{}
\makeatother
\setcopyright{rightsretained}
\usepackage{booktabs}
\usepackage{courier}
\usepackage{textcomp}

\usepackage{mdframed}
\setcopyright{none}
\setlist[itemize]{leftmargin=7mm}
\begin{document}

\title{\vspace{-2mm}Two Sides of the Same Coin: Software Developers' Perceptions of Task Switching and Task Interruption}

\author{Zahra Shakeri Hossein Abad$^{\textasteriskcentered}$, Mohammad Noaeen$^{\dagger}$, Didar Zowghi$^{\ddagger}$ , Behrouz H. Far$^{\dagger}$, Ken Barker$^{\textasteriskcentered}$}
\affiliation{%
\institution{$^{\textasteriskcentered}$ Department of Computer Science, University of Calgary, Canada, \{zshakeri, kbarker\}@ucalgary.ca}
 \institution{$^{\dagger}$ {\normalsize Department of Electrical and Computer Engineering}, University of Calgary, Canada, \{mohammad.noaeen, far\}@ucalgary.ca}
  \institution{$^{\ddagger}$ Faculty of Engineering and IT, University of Technology Sydney, Australia, didar.zowghi@uts.edu.au}
}

%
%

\begin{abstract}
In the constantly evolving world of software development, switching back and forth between tasks has become the norm. While task switching often allows developers to perform tasks effectively and may increase creativity via the flexible pathway, there are also consequences to frequent task-switching. For high-momentum tasks like software development, "flow", the highly productive state of concentration, is paramount. Each switch distracts the developers' flow, requiring them to switch mental state and an additional immersion period to get back into the flow. However, the wasted time due to time fragmentation caused by task switching is largely invisible and unnoticed by developers and managers. We conducted a survey with 141 software developers to investigate their perceptions of differences between task switching and task interruption and to explore whether they perceive task switchings as disruptive as interruptions. We found that practitioners perceive considerable similarities between the disruptiveness of task switching (either planned or unplanned) and random interruptions. The high level of cognitive cost and low performance are the main consequences of task switching articulated by our respondents. Our findings broaden the understanding of flow change among software practitioners in terms of the characteristics and categories of disruptive switches as well as the consequences of interruptions caused by daily\cm{stand-up} meetings. \cm{The results of this study provide valuable insights for practitioners about task switching and interruptions.}
\end{abstract}

%
%



\acmConference[EASE'18]{22nd International Conference on Evaluation and Assessment in Software Engineering 2018}{June 28--29, 2018}{Christchurch, New Zealand}
\acmBooktitle{EASE'18: 22nd International Conference on Evaluation and Assessment in Software Engineering 2018, June 28--29, 2018, Christchurch, New Zealand}
\acmDOI{-- } \acmISBN{ --}

\maketitle
\input{EASE-I2I-Body}

\vspace{-2mm}
\bibliographystyle{ACM-Reference-Format}
\bibliography{sample-bibliography} 

\end{document}

%% file: EASE-I2I-Body.tex
\section{Introduction}

Frequently, software development organizations are approving more projects to do than they have resources available to perform them. Thus, they need to assign software developers to multiple projects simultaneously to be able to handle all of the on-going projects. While this might seem a great solution, it overlooks the cost of "task switching" and time fragmentation caused by these switches. DeMarco and Lister \cite{Peopleware} define "flow" as a must-have for high-momentum tasks with a high level of cognitive demand involvement such as software development tasks. They also argue that once developers are locked in their current task, for each switch they need at least 15 minutes of concentration to be able to get back into the flow of the primary task. During this immersion period, they are very vulnerable to interruptions and environmental noise and are not really doing work \cite{Peopleware}. Parnin and Rugaber~\cite{Parnin} studied 10,000 programming sessions from 86 programmers and found that a developer's work, in a typical day, is cleaved into many short fragments, with a significant amount of time (i.e. 15-30 minutes) required to attain the flow before resuming switched tasks.

However, as the cost of fragmented time and frequent task switching is largely invisible, software practitioners, particularly in management positions are generally thought to be unconcerned about these costs and believe that as long as a flow change (e.g. task switchings, daily meetings) is pre-planned and necessary, it does not have any negative impact on developers' performance and productivity \cite{Peopleware}. To further investigate this perception, in this paper we gather data about software practitioners' perceptions of task switchings (either planned or unplanned) and interruptions through a survey of 141 professional software developers. Our analysis of the survey responses found that developers perceive that "task switching" and "interruptions" have similar consequences on their productivity and the cognitive cost caused by frequent task switching negatively impacts their performance in terms of increasing their error rate and the required time for completing the switched tasks. 

Moreover, as motivated by our industry partner and as previous studies have revealed that "daily meetings" are yet another form of interruption in workplace \cite{Daily,Daily1}, we were interested to get software practitioners' views of the extent of this interruption. \cm{Thus, we also studied the effects of daily meetings such as stand-up meetings on developers' productivity.} From this data, we found that the majority of developers perceive these meetings to take up more time than other interruptions and that they negatively impact their performance. Moreover, from this data, we gained some insight into the efficient timing of these meetings to help developers better manage their daily work flow. 

\cm{In the remainder of this paper, we first address needed concepts and terminologies and review the related work, in Section \ref{sec:RW}. Our research method, including research questions, survey design, and analysis is discussed in Section \ref{sec:Method}. We report and discuss the results of our study in Section \ref{sec:Results}, \cm{Limitations of this study and further discussion of our key findings are presented in Section \ref{sec:Discussion}}followed by conclusion in Section \ref{sec:Conclusion}.}

\section{Background and related work}
\label{sec:RW}

In this section, we describe basic concepts required to interpret the data obtained from survey respondents, followed by an overview of the related work.

\vspace{-2mm}
\subsection{Terminology}

{\em Task switching} is commonly considered as the act of toggling between separate tasks with different mental sets, attending to each independently \cite{Toggle}. A {\em task interruption} can be defined as any event that briefly shifts the attention of the subject from the on-going task towards some secondary external events \cite{Definition}.

Altmann and Trafton \cite{Memory} have posited a theory of "memory for goals" (called ACT-R) which explains the process of task switching in terms of general memory mechanisms of activation and associative priming. The central idea of this theory is that when people switch to a new task, the "goal" (i.e. what the current task is) of the new task and its associated "problem state" (i.e. the information required for performing this task \cite{Mind}) must be strengthened in memory, to the point where their activation rises above existing goals. Goals and associated information for achieving these goals are maintained in distinct areas of the brain. 
\begin{figure}
\centering
\includegraphics[scale=0.57]{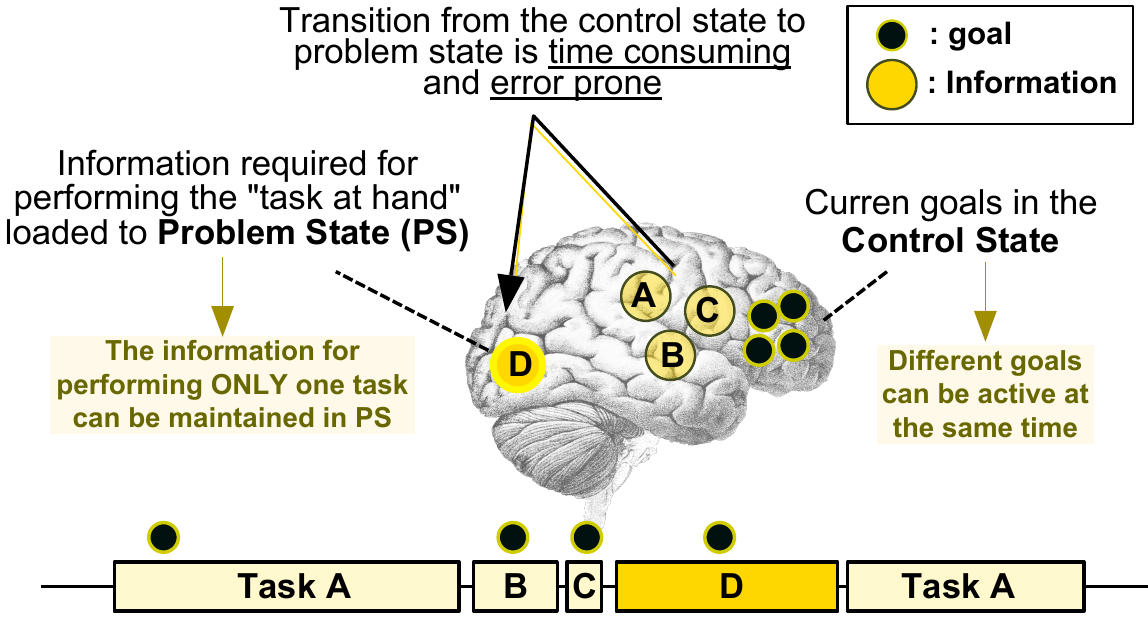}
\vspace{-3mm}
\caption{{Clarification of the concepts of "task switching", "problem state" and "goal memory". \copyright Shakeri H.A et al.}}
\label{fig:IBrain}
\vspace{-5mm}
\end{figure}
As illustrated in Figure \ref{fig:IBrain}, following the ACT-R theory, while human's cognition allows multiple goals to be active at the same time, only one problem state can be maintained at a time. Figure \ref{fig:IBrain} shows an example of a sequential task switching and the general memory mechanism of this process. Once a developer switches from Task $A$ to Task $B$, the problem state (required information for performing this task) of Task $A$  will be faded away (light yellow in this figure) and will be transferred from the "working memory" to the "declarative memory". The same process will happen to nested switches: $B \rightarrow C \rightarrow D$. When the developer is working on Task $D$, the problem state maintains only the information required for performing this task and the problem states associated with tasks $A$, $B$, and $C$ have to be swapped out when there is a switch between tasks. At the same time, the control state containing goals and information will decay according to the memory mechanisms incorporated into ACT-R \cite{Memory}. Thus, the required information for resuming a switched task that has been suspended and has fallen out of the active use will require more time and effort to recall. This process is not always successful; errors may occur when restoring the mental process back to the primary task.

\subsection{Related Work}
{In this section, we discuss related work that investigated the problem of task switching and interruptions in software development projects and evaluated their findings using industrial datasets. }

Abad et al. \cite{RE17, SERIP} conducted a set of experiments to evaluate the disruptiveness of task switching and interruptions in Requirements Engineering (RE) tasks to identify factors which make these interruptions more disruptive. They found that RE tasks are more vulnerable to the consequences of interruptions and several factors such as the source of the interruption (i.e. self/external), the granularity of the interrupted task, and the temporal aspects of the switching process (e.g. task's progress status) may impact the disruptiveness. Vasilescu et al. \cite{ICSE} studied the effects of task switchings on developers' productivity. They found that the {\it rate} and {\it the number of projects} in a task switching process are influential factors on developers' productivity. They also surveyed developers to understand the main reasons for, and perceptions of, multitaskings in software development. Participants of this study described the interrelationships and dependencies between projects as the most common reason for their task switchings.

Regarding resumption strategies, Parnin and Rugaber \cite{Parnin} conducted an analysis on 86 programmers to understand the various strategies and coping mechanisms that developers need to manage interrupted programming tasks. They found that only a small percentage of interrupted programming tasks were resumed in less than a minute. Chong and Siino \cite{Pair} compared interruption patterns among paired and solo programmers. Their study indicates significant differences between the pair programmers and solo programmers in terms of the length, type, time, context and strategies for handling task interruptions. They proposed that as a substantial number of interruptions are self-initiated, working in pairs may have potential support for interruption handling. Likewise, Stray et al. \citep{Daily} conducted a grounded theory study of 12 agile teams to explore developers' perceptions of daily stand-up meetings. They found that one of the prominent negative attitudes toward these meetings is that they are considered as an interruption to daily tasks and often occupy too much time relative to the gains from the meetings. While the past research provided a wealth of insight on task switching and interruptions, we could not find any study that investigated practitioners' perceptions of "task switching" and "interruption" as two distinct concepts.

\section{Research Design}
\label{sec:Method}
\cm{This section presents the research questions studied and outlines the research method we followed to implement our study.
\subsection{Goals and Research Question (RQs)}
In this study, we aim to investigate whether "task interruption" is a concept with a shared meaning and whether there is consensus on considering "task switching" as a form of interruptions. Also, as motivated by our industrial partners, we aim to study practitioners' perceptions of the disruptiveness of daily meeting. }

\cm{To gain a broad sense of what "task interruption" and " task switching" mean to software developer and whether there is any consensus on considering "task switching" as disruptive and interruptions, we raised RQ1 and RQ2.} In this study, we aim to address the following research questions: 
\begin{description}
\item [RQ1-] {Do software practitioners have a shared understanding of what "
task interruption" means?}

\item[RQ2-] {Do software practitioners consider "task switching" as disruptive as task interruptions? }

\item[RQ3-]{Do software practitioners consider switching to "daily stand-up meetings" as disruptive as a task interruption?}
\end{description}

\subsection{Survey Design and Participants}
To address our RQs, we designed and implemented an online survey, using Survey Monkey\cm{~\footnote{ http://www.surveymonkey.com [The survey questions and collected data will be available as an open data after receiving the notification]}}, to investigate software developers' perceptions and experiences of the disruptiveness of task interruptions and task switchings. 
The survey contained 15 questions including multiple choice, Likert scale, and open-ended questions: 5 on the practitioners' background and experience, 4 on perceptions of task switching and task interruptions, 3 on perceptions of the disruptiveness of daily meetings, and 3 on the management of task interruptions in a working environment. 

We sent the online survey to professional software developers working at eight software development companies of various sizes (e.g. Microsoft, Tableau Software, Ericsson, Bosch, Cisco, and CMG). To incentivize participants for high-quality data, we held a raffle for the on-line participants to win two \$50 Amazon gift certificates. 

We collected 141 valid answers from 10 countries, the highest population coming from the United States and Hungary with 48\% and 21\% of participants, respectively. Of 141 respondents, 80\% were male and 20\% were female, 106 (75\%) reported the size of their company greater than 1000 employees, 12 (9\%) between 100 and 1000, 23 (16\%) less than 100. The average professional software development experience per participant was 10.9 (range 1 to 40) years. The primary work
area of all participants was development: 93 (66\%) listed their job as a programmer, 21 (15\%) as a software architect, 17 (12\%) as a tester, 7 (5\%) as a project manager and 3 (2\%) as a requirements engineer. On an average day, 62 (44\%) contribute to one project, 38 (27\%) to two, 23 (16\%) to three, and 18 (13\%) to four or more projects. 
\begin{figure}
\centering
\vspace{-5mm}
\includegraphics[scale=0.62]{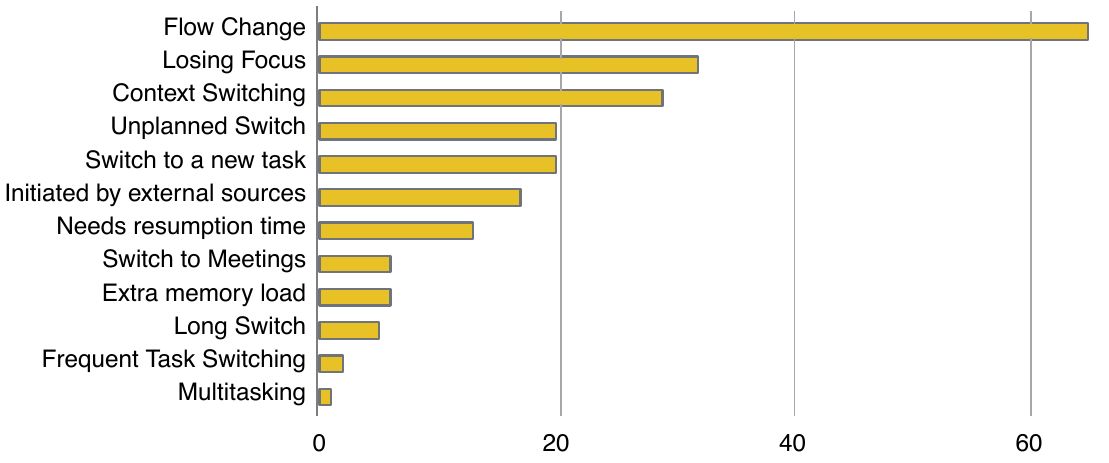}
\vspace{-3mm}
\caption{Coding frequency for open-ended question about the concept of ``task interruption''}
\vspace{-5.5mm}
\label{fig:RQ1}
\end{figure}

\subsection{Analysis}
For the open-ended questions, we iterated through these questions using open coding, axial coding, and selective coding (i.e. the grounded theory methods \cite{GT}). To implement this process and to coalesce all
of the references, we used NVivo \cite{Nvivo}, a qualitative analysis software package to code and analyze qualitative data. Additionally, to assess the correlation between participants' responses to survey questions, we used Spearman's rank test and considered $|\rho|\geq 0.50$ with {\em p-value}$<0.05$ as a strong significant correlation coefficient. Following the suggestion by Kitchenham et al. \cite{Density}, to compare the distribution of responses between different categories of respondents, we used kernel density plot instead of boxplots and used {\em Skew} and {\em Kurtosis} scores \cite{Kernel} to interpret these plots. 

\section{Results and Discussion}
\label{sec:Results}
In this section, we detail and discuss the results of our investigation related to each research question (RQ1-3).

\subsection{RQ1- Defining Task Interruption}

To answer this RQ, we asked respondents three open-ended questions to probe whether there was a shared understanding of what a "task interruption" is. The first two authors coded each response with at least one and at most five codes from a list of codes extracted during the first iteration of the coding process.  

"Flow change" and "losing focus" (Figure \ref{fig:RQ1}) are the two most commonly occurring codes on the open-ended question about the concept of task interruption. 80 (57\%) stated that a task interruption would be anything that breaks the flow of their thought, and distracts them from the current work. 40 (28\%) defined task interruption as any event that disrupts their focus on the "task at hand" and prevents them from working steadily on a single (or a set of) problem(s) in code: {\em ``A task interruption would be anything that takes my focus away from accomplishing the task goal. Particularly when I'm at a point in the task that I've cached a sufficient amount of information about the current state of the problem at hand in my memory. The interruption usually resets my cache and I need to start the task a few steps back from where I left off.''} Switching to a different task with a different context (context switching) and averting attention to another task (new task) are the next most frequent characteristics we {noted} for task interruption (context switching: 36 [26\%]; new task: 25 [18\%]): {\em ``Task interruption for me is something that requires context switching, that takes time and diverts my attention from the task I was working on, leading to an emptying and reloading of my brain's RAM''.} This is consistent with the general memory mechanism of task switching we illustrated in Figure \ref{fig:IBrain}, following the ACT-R and Memory-of-Goals theories \cite{Memory,Mind}.

25 (18\%) of respondents defined a task interruption as a spontaneous change in their work context, any unplanned activity which requires changing the thinking from the current task to another, such as coworkers coming to ask a question, instant messaging that requires a response, emails, or meetings. Apart from the context or the source of the interruption, some respondents characterized a task interruption as an event that requires restoring the mental process back to a given task: {\em ``When I have to stop doing something, and when I come back to it I have to spend at least a minute re-acquainting myself to the work''}.
\begin{figure}
\centering
\vspace{-5mm}
\includegraphics[scale=0.45]{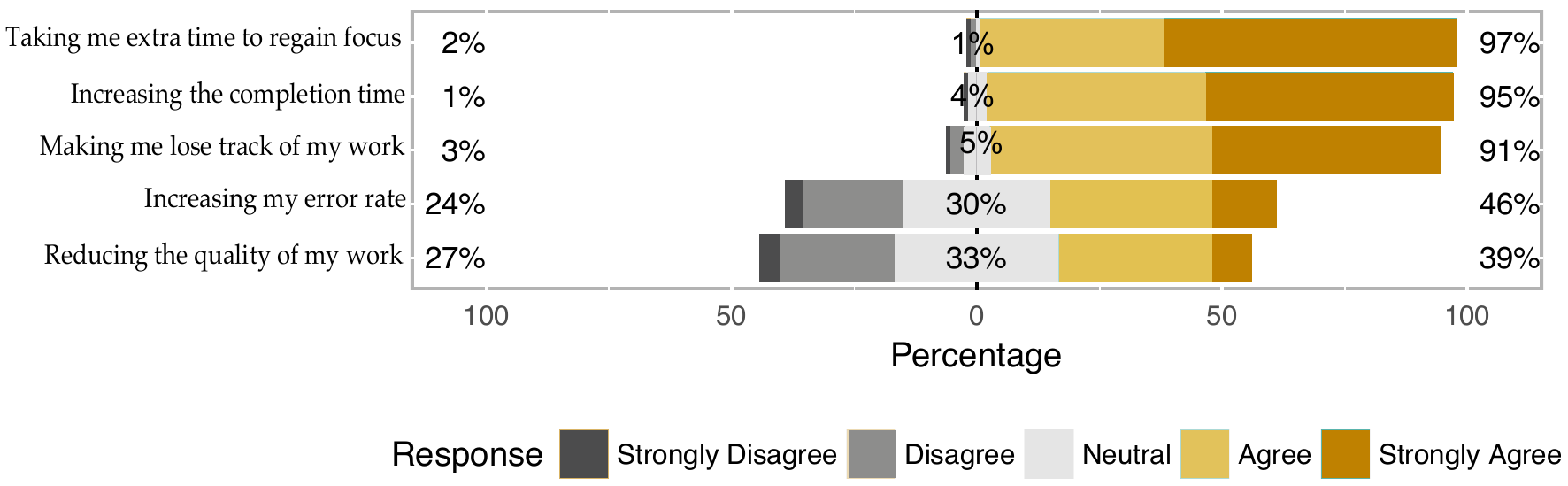}
\vspace{-5mm}
\caption{RQ1- Perceived consequences of task interruption}
\label{fig:Impact}
\vspace{-3mm}
\end{figure}
\begin{figure}
\centering
\includegraphics[scale=0.61]{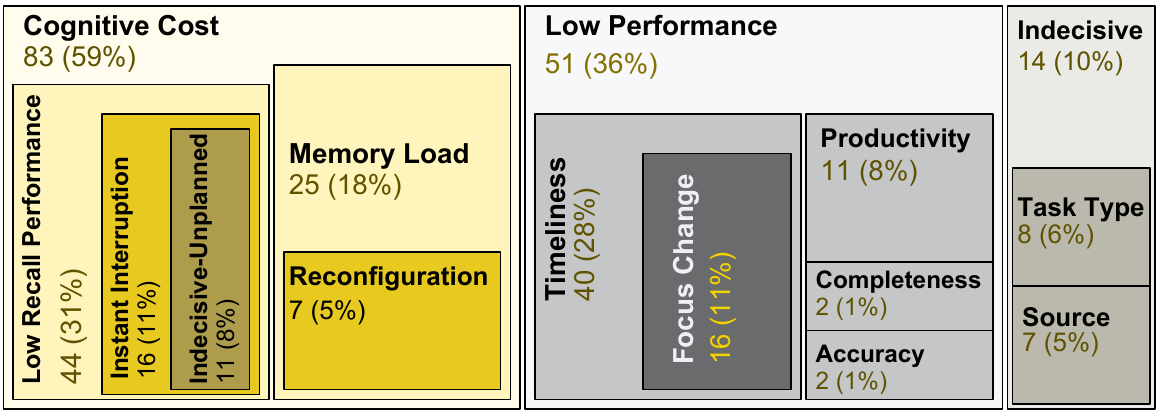}
\vspace{-3.5mm}
\caption{RQ2- Main reasons provided by respondents for considering task switching as interruption.}
\vspace{-5mm}
\label{fig:Nvivo}
\end{figure}
Of 141 responses to the statement ``A disruptive task interruption negatively impacts my performance after resuming the primary task by ...'', 97\% agreed or strongly agreed that ``the need for extra time to regain the focus'', 95\% that ``the need for extra time to complete the task'', and 91\% that ``losing the track of work'' are the consequences of interruptions on their performance (Figure \ref{fig:Impact}). 
Comparing to the open-ended question on the concept of task interruption, this suggests that there is a widespread agreement on {characteristics} of task interruption among respondents, including "flow change", "lose focus", and "timeliness".  Increasing the error rate and reducing the work quality are next in the list with each of them being 46\% and 39\% of agreement, respectively.

\subsection{RQ2-Task Interruption vs Task Switching}

When asked the participants ``Do you consider task switchings as a type of interruption?'', 115 (81\%) participants answered "Yes". We categorized the extracted concepts (during the coding process) from the responses to this question based on their similarities and identified the following main characteristics that make "task switchings" as disruptive as "task interruptions":

\subsubsection{Cognitive Cost}
83 (59\%) of the participants stated that task switchings, no matter if they are planned or ad-hoc, are a type of interruption because they cause an extra cognitive load: 
 {\em ``Definitely it is an interruption. Focusing on one task and finishing it is by far better than trying to fix everything at the same time. Focus/Concentration goes out of the window when trying to fix several different problems at the same time, be it in Software Development or in the kitchen while cooking several meals, eventually, something is bound to burn''}.
As illustrated in Figure \ref{fig:Nvivo}, we categorized the codes in the "cognitive cost" category based on their similarities and identified two main reasons of cognitive costs caused by task switching and were mentioned by our participants:

{\bf Low Recall Performance---} 44 (31\%) participants stated that switching to a new task would make their short-term memory need to "forget" about the interrupted task's context, so they can focus on the new information required to perform the new task. Thus, collecting thoughts to get to where they left off can be disruptive and time-consuming. For example, one participant stated: {\em ``When working on a task, I tend to become cognitively involved in that one task. So at that time in the moment, I have a deep understanding of the task at hand. But when I am required to task switch I lose that intimacy with the task I am working on. This hurts because when I come back to that task I spend some time getting back into that state of deep understanding''}.
16 (11\%) of the participants stated that task switching negatively impacts the recall performance if the switching request is unplanned (spontaneous) and happens randomly and so often (e.g. having to handle an incoming priority question, a support request from a tester)\cm{. This way, they cannot (mentally/physically) save the progress of the current task (e.g. leaving a note in the code with what they were going to do next), as well as the mental state which will be required for resuming the primary task later}, as in: {\em``Task switching requires some sort of wrap up on the current task and switching focus on a different topic that requires proper attention. If this occurs in an ad-hoc way, it can be considered as a type of interruption''.}

{\bf Memory Load---} Extra memory load was perceived by 25 (18\%) of the respondents as a source of cognitive cost which makes task switchings as disruptive as task interruptions: {\em ``If I turn my attention away from my code, even for a few seconds, I'll need a significant amount of time to get back to focus''.} Among these respondents, 7 (5\%) commented on the cognitive cost of the resumption process. For example: {\em ``It takes time to build up the context for any given task. To shelve edited files and re-build the code-base for the given task takes time and effort. There is an upfront cost to starting any task''}.

\subsubsection{Low Performance}
In addition to cognitive cost, as a factor which negatively impacts the disruptiveness of task switchings, 51 (36\%) respondents stated that "low performance" caused by losing focus is a consequence of task switchings, as in: {\em ``since most tasks differ in their context, switching between tasks decreases performance by forcing loading and unloading of context per switch''}. We categorized responses in this category to the following main sub-categories which cause a low performance after resuming the switched task:

{\bf Timeliness---}
40 (28\%) of the respondents stated that task switching is time-consuming and often does not increase performance to the point of justifying the time cost involved in the switching process: {\em ``There's a good amount of ramp-up time in each switch. It takes approximately 5-20 minutes to get into the flow state on the task at hand every time there is a switch''}. 16 (40\%) of these respondents (11\% of total) stated that task switching is an interruption on their thoughts (from one task to another) which causes them to lose focus and reduces their overall performance: {\em``I like to focus on what I'm doing until it's done. If I task switch, it causes me to interrupt my workflow and turns my attention away from my code, even for a few seconds, I'll need a significant amount of time to get back to focus''}.

{\bf Productivity---} 11(8\%) of the participants stated that they are more productive if they are doing only one thing at a time: {\em ``I find myself to be most productive and creative when I can be "in the groove" of a single piece of work for more than a day.''}

{\bf Completeness and Accuracy---} For each of completeness and accuracy categories we received two responses which indicate that task switching might negatively impact the completeness of the primary task: {\em ``At times the interruption makes me forget to go back to the previous task and finish it''}, or might increase the error rate: {\em ``When I have tried to perform two tasks concurrently, switching between them, I find that I make fairly substantial mistakes because I confuse the details of one task with the details of another \cm{So, if I need to frequently switch tasks, then I MUST use external memory aids like note-taking and diagramming software to capture EVERY detail}''}.

\vspace{-2mm}
\subsubsection{Indecisive}
Among respondents who answered "Yes" to our question about considering task switchings as disruptive as task interruptions, 14 (10\% \cm{,10\% of total}) were indecisive when it came to question about their choice. 8 (6\%) stated that the type and the granularity of the on-going and new tasks impact their answer to this question: {\em ``Depends on the type of task. If the task is totally new then it bothers otherwise (like a switch to a bug fix issue) it is not a big deal.''} Moreover, 7 (5\%\cm{, 5\% of total}) respondents indicated that the source of the task switching (i.e. initiated by themselves, or by external events  such as a production issue or helping someone track down a bug) impacts the disruptiveness of the switch: {\em ``If I'm been asked to switch tasks, it is an interruption. If I have made the decision myself for reasons such as taking a break from a task or wanting to tackle an easier/harder task first, I don't consider it an interruption''}.

\subsection{RQ3- Daily Stand-up Meetings}
To explore developers' perceptions of the disruptiveness of daily meetings, we asked them about the impact of switching to these meetings on their primary task's performance after the resumption process. Moreover, we asked them to identify a time when they feel the meetings are less disruptive to their development tasks.

\subsubsection{Disruptiveness of Daily Meetings}
We asked respondents to complete the following
sentence stem:``If I switch my current ongoing task to a daily team meeting, this interruption has a $\ldots$ impact on my primary task's performance after resuming this task''. \cm{As illustrated in Figure \ref{fig:Meeting}a, }64 (45\%) stated that switching to daily meetings has a negative or very negative impact on their performance after resuming the primary task. These respondents believed that discussions in these meetings are more general and unrelated to specific tasks which make daily meetings not informative enough for most people: 
{\em ``It's an interruption like any others unless I needed help and got it in the meeting. Time after meetings is usually my least productive''}. Moreover, based on the results of our coding process, 5 respondents stated that not only do these meetings take up their time, they stop them from engaging in meaningful work schedule before and after: {\em ``Say there is a meeting at 11, and 1 pm and you usually eat lunch at 12. You simply will not be able to start a task that needs 90 minutes of focus from 10am-2pm. And rarely can you estimate that. I tend to think of "real work" in 2-hour blocks and anything less will just be "busy work" like responding to emails and helping others''}. This is consistent with what Stray et al. \cite{Daily} and Solingen et al. \cite{Daily1} found: developers found the time of daily meetings disruptive (as it results in an undesirable long resumption lag). 

On the other hand, 32 (23\%) respondents acknowledged that daily meetings have a (very) positive impact on their performance after resuming the primary task. Getting context on where the rest of the team is at, getting help to unblock each other, and clarifying tasks and their priority are the main reasons provided by these respondents: {\em ``while they might negatively impact my task performance, the utility of stand-up outweighs that. Regular meetings keep everyone on the same page, so what may take 2 min in the stand-up could take 15+ in a chat''.} Moreover, 45 (31\%) participants believed that daily meetings are something planned and expected, and have no impact on their performance: {\em ``If it's a regular meeting and it's a short one, I don't expect an impact. My brain knows it's coming''}.

\begin{figure}
\centering
\includegraphics[scale=0.81]{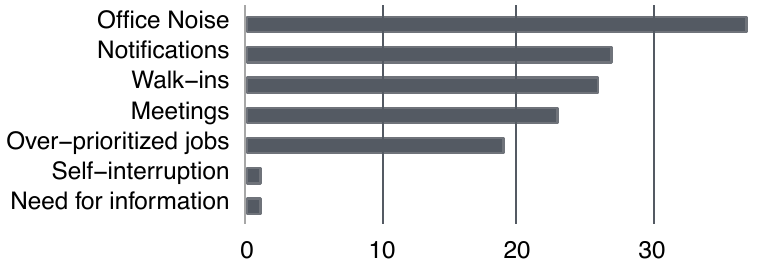}
\vspace{-4mm}
\caption{Coding frequency for open-ended question about the most disruptive interruption}
\vspace{-6mm}
\label{fig:Discussion}
\end{figure}

\subsubsection{A Productive Meeting Time}
To further investigate the association between the time and the disruptive impact of daily meetings, and as requested by our industrial partner after reporting the results of our second pilot study, we asked respondents about temporal aspects of the daily meetings in their company. When asked ``Is there any specific (pre-defined) time slot for meetings in your company?'', 99 (70\%) responded "No", 42 (30\%) answered "Yes" {[}30 (21\%): ``Yes, usually before noon''; 12 (9\%): ``Yes, usually afternoon''{]}. Moreover, to check the correlation between participants' response to this question and to the question about the disruptiveness of switching to daily meetings, we used the {\em Spearman's correlation rank test} and defined $|\rho| \geq 0.50$ as a strong correlation coefficient. The results of this test show that there is a week correlation between the perceived disruptiveness of daily meetings ranked by participants and the schedule of meetings in the company they work for (rho=0.2, {\em p-value}= 0.04). \cm{This implies that respondents who stated switching to daily meetings negatively impact their performance on the primary task are working in a company without any pre-defined schedule for daily meetings.}

Moreover, we asked participants which time for scheduling daily meetings has the least negative impact on their performance among "morning, before starting daily tasks", "anytime in the morning", "right after noon", and "afternoon, as the last daily task". The majority of respondents (100, 71\%) perceived "morning meetings" less disruptive to their daily tasks, among which 62 (44\%) respondents stated that it might be better to get the meeting done before starting any tasks than to interrupt them to have a meeting that may or may not be productive \cm{(see Figure \ref{fig:Meeting}b)}:  {\em ``I tend to do my best work in the late morning and early afternoon. Since our team stand-up is in the morning, it's helpful to know what I need to do and then have a long uninterrupted period of time to do it in''}. Scheduling the meetings for "right after noon" comes next with 22 (15\%) responses. These respondents believe that there already is a disruption at noon (eating lunch), thus, by scheduling the meetings around noon they can have a nice contiguous block of disruptions rather than have the meetings all spaced out:  {\em``I try to batch my meetings together to make long spans of uninterrupted time. Lunchtime is the best time to synchronize with a common interruption; morning is harder because not everyone gets in at the same time''}. Scheduling the meetings as the last daily task comes last with 20 (15\%) of responses. Seven respondents in this category believed that meetings at the later time of a day give developers more time to prepare and finish final touches on their work before they have to discuss them. However, the rest of respondents perceived this time help decrease the disruptiveness of daily meetings: {\em ``Meetings are mind-numbing and should be done when no further thinking is required afterward. Mornings tend to be more productive for me to get focused work done''}.

\cm{
\subsection{Threats to Validity}
We followed and adapted the survey guidelines in software engineering provided by Moll{\'e}ri et al. \cite{Survey} to mitigate the limitations of survey research. To evaluate the reliability of our coding process (for analyzing the open-ended questions), we used the Cohen's Kappa statistics to measure the degree of agreement and consistency between the extracted themes by the first two authors who were involved in the coding process. The calculated Kappa value was 0.83, which shows significant agreement according to Landis and Koch \cite{Kappa}. In regard to survey questions, we pilot tested the survey questions in our two previous pilot studies to mitigate the risk of misunderstanding questions. However, as the questions still require participants' interpretation, we added a comment box for each question and asked respondents to clarify their response or further discuss other aspects of the question. 

Our survey respondents might not be representative of the entire software development community, our survey result might thus suffer from a lack of generalizability. We mitigated this threat by distributing our survey to a large number of potential respondents with different levels of software development experience and from various countries (e.g. Germany, Netherland, Sweden, Hungary, USA, New Zealand, and Canada).}

\section{Practitioners' Corner}

\subsection{Lessons Learned}
Following important lessons are aggregated:

\begin{itemize}
\item Regardless of the interruption characteristics and type, as stated by our participants, {\em flow change} and {\em losing focus} are the main reasons which make interruptions costly and harmful to the overall performance of the primary task. This suggests that methods for securing and rebuilding developers' focus can provide value.

\item The majority (81\%) of respondents of our study perceive that "task switching" is as disruptive as "interruption" because of the {\em cognitive cost} (e.g. low recall performance,  memory load) and {\em low performance} (e.g. timeliness and the negative impact of task switching on productivity, completeness, and accuracy) caused by task switchings.

\item Respondents who perceived that switching to daily meetings negatively impact their performance on the primary task are working in a company without any pre-defined schedule for daily meetings. Losing the flow of the on-going task to attend a meeting appears to be a reason contributing to daily meetings being perceived as disruptive: {\em``Meetings are the \#1 productivity killer, especially the daily stand-ups. These meetings often sneak up on me since I'm so engaged in my current task, which means that I most likely did not stop at a good place (i.e. I stop in the middle of a subtask)''}.

\item The timing and content of daily meetings are perceived to be influential on the disruptiveness of these meetings. Frequent meetings and large meetings that have more to do with process and organizational changes tend to be more disruptive than planned and short meetings. The results of our analysis suggest that it might be better to get the meeting done before starting any tasks than to interrupt them to have a meeting that may or may not be productive.
\end{itemize}

\subsection{Recommendations}

{\bf Sources of Interruptions: }When we asked respondents about the most disruptive sources of interruptions in their working environment, as illustrated in Figure \ref{fig:Discussion}, the majority of respondents stated that "office noise" such as overhearing co-workers' phone calls in an open office, or the random chatter of colleagues on the other side of the office is the  most disruptive source of distraction to their work: {\em ``Just wish we could find a way to manage "focus time" more explicitly. We work in an open office so that everyone knows about everyone's progress, but this also means that someone is sitting facing me. When she looks up it feels like looking at me, which is kind of awkward ... too bad we don't have an "I am focusing" signal''}. This is in conformance with Glass's finding: {``The working environment has a profound impact on productivity and product quality''} \cite{Peopleware}. \cm{He discusses that in addition to the profound effect that individual differences between people can make on the productivity, there are some other factors that must be taken into account.} The other most commonly occurring codes on respondents' answers to this question, as illustrated in Figure \ref{fig:Discussion}, are notifications, walk-ins, meetings, over-prioritized jobs, self-interruption, and need for information.

{\bf Notifications:} To manage the disruptiveness of notifications, we propose that it may be less disruptive in some situations (e.g. where a developer just initiated a task) to delay receiving/addressing notifications, as stated by one of our participants: {\em ``when I join a team, I turn off all notifications, and then selectively turn notifications back on as I find they reflect information I need in a more timely fashion. While this does degrade my performance a bit, I do this for cases where I decide that degradation is worth the information--- and I also block out all notifications when I'm super focused''.}

\begin{figure}
\centering
\includegraphics[scale=0.53]{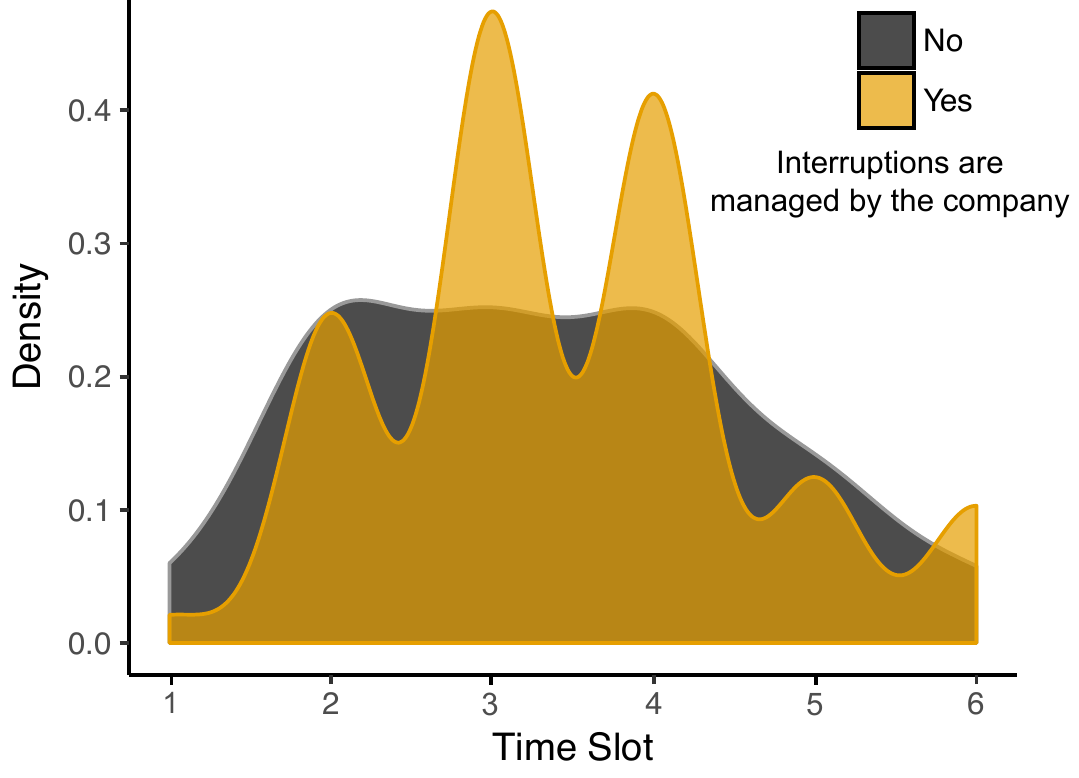}
\vspace{-5.5mm}
\caption{\footnotesize Answers to question: ``On a typical working day, how often you switch your software development tasks?'' [1]: Less than every 10 minutes, [2]: Every 10-30 minutes, [3]: Every 30-60 minutes, [4]: Every 1-2 hours, [5]: Every 2-4 hours, [6]: Others.}
\vspace{-5mm}
\label{fig:Compare}
\end{figure}

{\bf Frequency of Task Switching:} To consider environmental factors such as the working environment and communication methods, before asking about the frequency of task switchings, we first asked participants whether their company use any technique to protect them against task switching and interruptions. 75 (53\%) responded "Yes". Having frequent Scrum meetings to focus on only planned items, using noise-canceling headphones, keeping the scope of the meetings minimal, defining {\em library hours} during the week, no-meeting Mondays, and improving build speeds are some example techniques mentioned by our respondents. Then, respondents were asked to define the frequency of their task switching. The density plots are shown in Figure \ref{fig:Compare} summarize and compare the distribution of answers to this question. Comparing the two density plots, it is clear that the shape of distribution of answers from developers who work in a company with no clearly defined strategies for managing interruptions is more homogeneous (Skew$_{no}$= 0.26, Skew$_{yes}$=0.43) and light-tailed (Kurtosis$_{no}$=-0.7, Kurtosis$_{yes}$=-0.2) across the range of possible scores compared to the "Yes" category, i.e. the concentration of values associated with the "Yes" category is higher towards the "every 30-60 minutes" and "every 1-2 hours" responses. One possible explanation for this could be that respondents in this category are better protected from random and spontaneous interruptions by their company. Considering the unproductive reimmersion time associated with each task switching or interruption, frequent task switching during a day can use up most of that day. 

{\bf Management:} 
As stated by our participants, paranoia interruptions or anything with some anxiety attached to it would be harder to overcome. For example, if a colleague comes in with an invitation to share a treat, it is way less interruptive that someone calling for a harsh issue, as in: {\em ``we had a manager who stack ranked primarily on the number of code reviews (CR) performed. Everyone was on a hair trigger for when CR requests came through, they would drop what they were doing and fight to be in the first four reviewers (the only ones that counted)''}. People have different degrees of tolerance for interruptions, urgent new tasks or intermittent interruptions can be detrimental to developers' productivity. Our results suggest setting expectations and boundaries for your team and yourself. The more you can reduce or plan around potential interruptions, the better.

\section{Conclusion}
\label{sec:Conclusion}
To investigate how developers perceive and distinguish the concepts of "task switching" and "task interruption", we conducted a survey with 141 professional software developers from across the world. The survey results show that while most
participants considered task interruptions and task switchings disruptive (due to the flow change, losing focus, context switching, etc), 79 (56\%) respondents reported that, in a typical day, they cannot be fully focused and frequently flip back and forth between different tasks (45 (32\%): less than every 30 minutes, 34 (24\%): every 30-60 minutes, RQ1). The results also show that practitioners perceive task switchings are as disruptive as spontaneous and random interruptions, regardless of the source and type of the switching (RQ2). The heavy cognitive load and the low performance caused by frequent task switching are the main reasons for this perception. In addition, we explored practitioners' perceptions of the disruptiveness of daily meetings as well as their perceptions of a productive meeting time. While some participants considered daily meetings useful in terms of clarifying daily tasks and gaining valuable input on how they should proceed, the majority of respondents perceived the impact of daily meetings to be (very) negative (RQ3).  \cm{Particularly, they stated that meetings that have more to do with process and organizational decisions tend to take them out of the head-space and are very disruptive.} The findings of this study are in line with our previous studies \cite{SERIP, RE17, Abad2018, Abad2017} on the disruptiveness of task switching and task interruptions.

\cm{Our future work directions include the following: (1) developing a tool to detect and to visualize task switchings and interruptions to help developers retrospecting on their own flow and productivity; (2) replicating our study in different companies with different development process models to investigate the effectiveness of existing software development methodologies in managing task switchings and interruptions. }

%% file: EASE-I2I-2018.bbl

\begin{thebibliography}{00}


\ifx \showCODEN    \undefined \def \showCODEN     #1{\unskip}     \fi
\ifx \showDOI      \undefined \def \showDOI       #1{#1}\fi
\ifx \showISBNx    \undefined \def \showISBNx     #1{\unskip}     \fi
\ifx \showISBNxiii \undefined \def \showISBNxiii  #1{\unskip}     \fi
\ifx \showISSN     \undefined \def \showISSN      #1{\unskip}     \fi
\ifx \showLCCN     \undefined \def \showLCCN      #1{\unskip}     \fi
\ifx \shownote     \undefined \def \shownote      #1{#1}          \fi
\ifx \showarticletitle \undefined \def \showarticletitle #1{#1}   \fi
\ifx \showURL      \undefined \def \showURL       {\relax}        \fi
\providecommand\bibfield[2]{#2}
\providecommand\bibinfo[2]{#2}
\providecommand\natexlab[1]{#1}
\providecommand\showeprint[2][]{arXiv:#2}

\bibitem[\protect\citeauthoryear{Abad, Karras, Schneider, Barker, and
  Bauer}{Abad et~al\mbox{.}}{2018}]%
        {Abad2018}
\bibfield{author}{\bibinfo{person}{Zahra Shakeri~Hossein Abad},
  \bibinfo{person}{Oliver Karras}, \bibinfo{person}{Kurt Schneider},
  \bibinfo{person}{Ken Barker}, {and} \bibinfo{person}{Mike Bauer}.}
  \bibinfo{year}{2018}\natexlab{}.
\newblock \showarticletitle{Task Interruption in Software Development Projects:
  What Makes some Interruptions More Disruptive than Others?}. In
  \bibinfo{booktitle}{{\em Proceedings of the 22nd International Conference on
  Evaluation and Assessment in Software Engineering}} {\em
  (\bibinfo{series}{EASE'18})}. \bibinfo{publisher}{ACM}.
\newblock


\bibitem[\protect\citeauthoryear{Abad, Ruhe, and Bauer}{Abad
  et~al\mbox{.}}{2017a}]%
        {RE17}
\bibfield{author}{\bibinfo{person}{Zahra Shakeri~Hossein Abad},
  \bibinfo{person}{Guenther Ruhe}, {and} \bibinfo{person}{Mike Bauer}.}
  \bibinfo{year}{2017}\natexlab{a}.
\newblock \showarticletitle{{Task Interruptions in Requirements Engineering:
  Reality versus Perceptions!}}. In \bibinfo{booktitle}{{\em Requirements
  Engineering Conference (RE), 2017 IEEE 25th International}}. IEEE,
  \bibinfo{pages}{6--15}.
\newblock


\bibitem[\protect\citeauthoryear{Abad, Ruhe, and Bauer}{Abad
  et~al\mbox{.}}{2017b}]%
        {SERIP}
\bibfield{author}{\bibinfo{person}{Zahra Shakeri~Hossein Abad},
  \bibinfo{person}{Guenther Ruhe}, {and} \bibinfo{person}{Mike Bauer}.}
  \bibinfo{year}{2017}\natexlab{b}.
\newblock \showarticletitle{{Understanding Task Interruptions in Service
  Oriented Software Development Projects: An Exploratory Study}}. In
  \bibinfo{booktitle}{{\em Proceedings of the 4th International Workshop on
  Software Engineering Research and Industrial Practice}} {\em
  (\bibinfo{series}{SER\&IP '17})}. \bibinfo{publisher}{IEEE Press},
  \bibinfo{pages}{34--40}.
\newblock


\bibitem[\protect\citeauthoryear{Abad, Shymka, Le, Hammad, and Ruhe}{Abad
  et~al\mbox{.}}{2017c}]%
        {Abad2017}
\bibfield{author}{\bibinfo{person}{Zahra Shakeri~Hossein Abad},
  \bibinfo{person}{Alex Shymka}, \bibinfo{person}{Jenny Le},
  \bibinfo{person}{Noor Hammad}, {and} \bibinfo{person}{Guenther Ruhe}.}
  \bibinfo{year}{2017}\natexlab{c}.
\newblock \showarticletitle{{A Visual Narrative Path from Switching to Resuming
  a Requirements Engineering Task}}. In \bibinfo{booktitle}{{\em Requirements
  Engineering Conference (RE), 2017 IEEE 25th International}}. IEEE,
  \bibinfo{pages}{442--447}.
\newblock


\bibitem[\protect\citeauthoryear{Adolph, Hall, and Kruchten}{Adolph
  et~al\mbox{.}}{2011}]%
        {GT}
\bibfield{author}{\bibinfo{person}{Steve Adolph}, \bibinfo{person}{Wendy Hall},
  {and} \bibinfo{person}{Philippe Kruchten}.} \bibinfo{year}{2011}\natexlab{}.
\newblock \showarticletitle{{Using grounded theory to study the experience of
  software development}}.
\newblock \bibinfo{journal}{{\em Empirical Software Engineering\/}}
  \bibinfo{volume}{16}, \bibinfo{number}{4} (\bibinfo{year}{2011}),
  \bibinfo{pages}{487--513}.
\newblock


\bibitem[\protect\citeauthoryear{Altmann and Trafton}{Altmann and
  Trafton}{2002}]%
        {Memory}
\bibfield{author}{\bibinfo{person}{Erik~M Altmann} {and}
  \bibinfo{person}{J.Gregory Trafton}.} \bibinfo{year}{2002}\natexlab{}.
\newblock \showarticletitle{{Memory for goals: an activation-based model}}.
\newblock \bibinfo{journal}{{\em Cognitive Science\/}} \bibinfo{volume}{26},
  \bibinfo{number}{1} (\bibinfo{year}{2002}), \bibinfo{pages}{39 -- 83}.
\newblock


\bibitem[\protect\citeauthoryear{Chisholm~et al.}{Chisholm~et al.}{2000}]%
        {Definition}
\bibfield{author}{\bibinfo{person}{Carey~D Chisholm~et al.}}
  \bibinfo{year}{2000}\natexlab{}.
\newblock \bibinfo{journal}{{\em Academic Emergency Medicine\/}}
  \bibinfo{volume}{7}, \bibinfo{number}{11} (\bibinfo{year}{2000}),
  \bibinfo{pages}{1239--1243}.
\newblock


\bibitem[\protect\citeauthoryear{Chong and Siino}{Chong and Siino}{2006}]%
        {Pair}
\bibfield{author}{\bibinfo{person}{Jan Chong} {and} \bibinfo{person}{Rosanne
  Siino}.} \bibinfo{year}{2006}\natexlab{}.
\newblock \showarticletitle{{Interruptions on Software Teams: A Comparison of
  Paired and Solo Programmers}}. In \bibinfo{booktitle}{{\em Proceedings of the
  2006 20th Anniversary Conference on Computer Supported Cooperative Work}}.
  \bibinfo{publisher}{ACM}, \bibinfo{pages}{29--38}.
\newblock


\bibitem[\protect\citeauthoryear{DeMarco and Lister}{DeMarco and
  Lister}{2013}]%
        {Peopleware}
\bibfield{author}{\bibinfo{person}{Tom DeMarco} {and} \bibinfo{person}{Tim
  Lister}.} \bibinfo{year}{2013}\natexlab{}.
\newblock \bibinfo{booktitle}{{\em {Peopleware: Productive Projects and
  Teams}}}.
\newblock \bibinfo{publisher}{Addison-Wesley}.
\newblock


\bibitem[\protect\citeauthoryear{et~al.}{et~al.}{1998}]%
        {Daily1}
\bibfield{author}{\bibinfo{person}{Van~Solingen et al.}}
  \bibinfo{year}{1998}\natexlab{}.
\newblock \showarticletitle{{Interrupts: Just a Minute Never Is}}.
\newblock \bibinfo{journal}{{\em IEEE software\/}} \bibinfo{volume}{15},
  \bibinfo{number}{5} (\bibinfo{year}{1998}), \bibinfo{pages}{97--103}.
\newblock


\bibitem[\protect\citeauthoryear{Gibbs}{Gibbs}{2002}]%
        {Nvivo}
\bibfield{author}{\bibinfo{person}{Graham~R Gibbs}.}
  \bibinfo{year}{2002}\natexlab{}.
\newblock \bibinfo{booktitle}{{\em Qualitative data analysis: Explorations with
  NVivo}}.
\newblock \bibinfo{publisher}{Open University}.
\newblock


\bibitem[\protect\citeauthoryear{Kitchenham~et al.}{Kitchenham~et al.}{2017}]%
        {Density}
\bibfield{author}{\bibinfo{person}{Barbara Kitchenham~et al.}}
  \bibinfo{year}{2017}\natexlab{}.
\newblock \showarticletitle{{Robust Statistical Methods for Empirical Software
  Engineering", journal="Empirical Software Engineering}}.
\newblock  \bibinfo{volume}{22}, \bibinfo{number}{2} (\bibinfo{year}{2017}),
  \bibinfo{pages}{579--630}.
\newblock


\bibitem[\protect\citeauthoryear{Parnin and Rugaber}{Parnin and
  Rugaber}{2011}]%
        {Parnin}
\bibfield{author}{\bibinfo{person}{Chris Parnin} {and} \bibinfo{person}{Spencer
  Rugaber}.} \bibinfo{year}{2011}\natexlab{}.
\newblock \showarticletitle{{{Resumption Strategies for Interrupted Programming
  Tasks}}}.
\newblock \bibinfo{journal}{{\em Software Quality Journal\/}}
  \bibinfo{volume}{19}, \bibinfo{number}{1} (\bibinfo{year}{2011}),
  \bibinfo{pages}{5--34}.
\newblock


\bibitem[\protect\citeauthoryear{Salvucci and Taatgen}{Salvucci and
  Taatgen}{2010}]%
        {Mind}
\bibfield{author}{\bibinfo{person}{Dario~D Salvucci} {and}
  \bibinfo{person}{Niels~A Taatgen}.} \bibinfo{year}{2010}\natexlab{}.
\newblock \bibinfo{booktitle}{{\em {The multitasking mind}}}.
\newblock \bibinfo{publisher}{Oxford University Press}.
\newblock


\bibitem[\protect\citeauthoryear{Scott}{Scott}{1979}]%
        {Kernel}
\bibfield{author}{\bibinfo{person}{David~W Scott}.}
  \bibinfo{year}{1979}\natexlab{}.
\newblock \showarticletitle{{On Optimal and Data-based Histograms}}.
\newblock \bibinfo{journal}{{\em Biometrika\/}} \bibinfo{volume}{66},
  \bibinfo{number}{3} (\bibinfo{year}{1979}), \bibinfo{pages}{605--610}.
\newblock


\bibitem[\protect\citeauthoryear{Stray, Sj{\o}berg, and Dyb{\aa}}{Stray
  et~al\mbox{.}}{2016}]%
        {Daily}
\bibfield{author}{\bibinfo{person}{Viktoria Stray}, \bibinfo{person}{Dag~IK
  Sj{\o}berg}, {and} \bibinfo{person}{Tore Dyb{\aa}}.}
  \bibinfo{year}{2016}\natexlab{}.
\newblock \showarticletitle{{The Daily Stand-up Meeting: A Grounded Theory
  Study}}.
\newblock \bibinfo{journal}{{\em Journal of Systems and Software\/}}
  \bibinfo{volume}{114} (\bibinfo{year}{2016}), \bibinfo{pages}{101 -- 124}.
\newblock


\bibitem[\protect\citeauthoryear{Terry, Mishra, and Roseth}{Terry
  et~al\mbox{.}}{2016}]%
        {Toggle}
\bibfield{author}{\bibinfo{person}{Colin~A. Terry}, \bibinfo{person}{Punya
  Mishra}, {and} \bibinfo{person}{Cary~J. Roseth}.}
  \bibinfo{year}{2016}\natexlab{}.
\newblock \showarticletitle{{Preference for multitasking, technological
  dependency, student metacognition, \& pervasive technology use: An
  experimental intervention}}.
\newblock \bibinfo{journal}{{\em Computers in Human Behavior\/}}
  \bibinfo{volume}{65} (\bibinfo{year}{2016}), \bibinfo{pages}{241 -- 251}.
\newblock


\bibitem[\protect\citeauthoryear{Vasilescu}{Vasilescu}{2016}]%
        {ICSE}
\bibfield{author}{\bibinfo{person}{Bogdan et~al. Vasilescu}.}
  \bibinfo{year}{2016}\natexlab{}.
\newblock \showarticletitle{{The Sky is Not the Limit: Multitasking Across
  GitHub Projects}}. In \bibinfo{booktitle}{{\em Proceedings of the 38th
  International Conference on Software Engineering}}.
  \bibinfo{pages}{994--1005}.
\newblock


\end{thebibliography}
